\documentclass[useAMS,usenatbib]{mn2e}
\usepackage{amssymb}
\usepackage{graphicx}

\newcommand{\object}{SDSS\,J080449.49+161624.8}
\newcommand{\obj}{SDSS\,J0804+1616}
\newcommand{\porb}{$P_\mathrm{orb}=44.5\pm0.1$\,min}
\newcommand{\gpm}[2]{$#1\pm#2$}
\newcommand{\sample}{1,523}
\newcommand{\sampleapprox}{1,500}
\newcommand{\vetoed}{65}
\newcommand{\done}{15\%}

\title[\obj: a peculiar new AM CVn star]{\object: A peculiar AM CVn star from a colour-selected sample of candidates}

\author[Roelofs et al.]{G.\,H.\,A.~Roelofs,$^{1}$\thanks{E-mail: groelofs@cfa.harvard.edu}
  P.\,J.~Groot,$^{2}$ D.~Steeghs,$^{3}$ A.~Rau,$^{4}$  E.~de Groot,$^2$ T.\,R.~Marsh,$^{3}$
  \newauthor{G.~Nelemans,$^{2}$ J. Liebert,$^{5}$ P.~Woudt$^{6}$}
  \\
  $^{1}$Harvard--Smithsonian Center for Astrophysics, 60 Garden Street,
  Cambridge, MA 02138, USA\\
  $^{2}$Department of Astrophysics, Radboud University
  Nijmegen, PO Box 9010, 6500 GL Nijmegen, The Netherlands\\
  $^3$Department of Physics, University of Warwick, Coventry CV4 7AL, UK\\
  $^4$Astronomy Department, California Institute of Technology, Pasadena, CA 91125, USA\\
  $^5$Steward Observatory, University of Arizona, Tucson, AZ 85721, USA\\
  $^6$Department of Astronomy, University of Cape Town, Rondebosch 7700, South Africa
}

\begin{document}

\date{Accepted ... Received \today}

\pagerange{\pageref{firstpage}--\pageref{lastpage}} \pubyear{2008}

\maketitle

\label{firstpage}

\begin{abstract}
We describe a spectroscopic survey designed to uncover an estimated $\sim$40 AM CVn stars hiding in the photometric database of the Sloan Digital Sky Survey (SDSS). We have constructed a relatively small sample of about \sampleapprox\ candidates based on a colour selection, which should contain the majority of all AM CVn binaries while remaining small enough that spectroscopic identification of the full sample is feasible.

We present the first new AM CVn star discovered using this strategy, \object, the ultracompact binary nature of which is demonstrated using high-time-resolution spectroscopy obtained at the Magellan telescopes at Las Campanas Observatory, Chile. A kinematic `S-wave' feature is observed on a period \porb, which we propose is the orbital period, although the present data cannot yet exclude its nearest daily aliases.

The new AM CVn star shows a peculiar spectrum of broad, single-peaked helium emission lines with unusually strong series of ionised helium, reminiscent of the (intermediate) polars among the hydrogen-rich Cataclysmic Variables. We speculate that \obj\ may be the first magnetic AM CVn star. The accreted material appears to be enriched in nitrogen, to N/O$\,\gtrsim\,$10 and N/C$\,>\,$10 by number, indicating CNO-cycle hydrogen burning, but no helium burning, in the prior evolution of the donor star.
\end{abstract}


\section{Introduction}
\label{introduction}
The AM CVn stars are a small, but growing class of ultracompact binaries
that consist of a white dwarf accretor and a \mbox{(semi-)}degenerate
helium-transferring donor star. They are characterized by their short orbital
periods (below the orbital period minimum for hydrogen-rich donors) and the lack of hydrogen in their
spectra. There are currently 22 members,
including the new system reported here and the two ultrashort-period candidates HM Cnc
(RX\,J0806+15, \citealt{israel}) and V407 Vul (RX\,J1914+24, \citealt{cropper}).
A recent overview is given by \citet{nelemans2005}.

Our understanding of the AM CVn stars has been increasing rapidly
in the last few years. This is due to more detailed observational studies
of known systems (e.g. \citealt{roelofsamcvn,roelofshst}) and more detailed physical models (e.g.\ \citealt{deloye,deloye2007,bildsten,bildstenIa,yungelson}), but in large part also
due to an increase in the number of objects known
(e.g.\ \citealt{ww2003aw,ww1427,roelofs,anderson,anderson08}). The Sloan Digital Sky Survey (SDSS;
\citealt{york}) has proven to be an important resource for the
discovery of new AM CVn stars, with six new systems so far. More importantly, since the SDSS sample is relatively well defined, those six systems provide information about the underlying Galactic population that is difficult to obtain from the rather random individual discoveries from the pre-SDSS era (Roelofs, Nelemans \& Groot 2007b).

We have shown in \citet{roelofspop} that the photometric database of the SDSS, Phase I (SDSS-I, or Data Release 5) should contain a total of about 50 AM CVn stars down to a magnitude of $g=21$ (but with substantial uncertainty of a factor 2--3 due to, in part, the small sample of six systems from the spectroscopic database).
We have started an
extensive observing program using an array of telescopes to uncover
this `hidden' population of AM CVn stars. The significance of this sample will
not only be to double or triple the total known population, but more
importantly to increase the number of systems that have been found in a
relatively well-understood way by an even larger factor, which will allow for a much more detailed comparison of their population with theoretical predictions.

The outline of this paper is as follows. In Section\ \ref{sec:sample} we discuss the total sample of AM CVn candidates we selected from
the SDSS, Data Release 6 (SDSS-DR6; \citealt{adelman}). In Section\ \ref{sec:observing} we discuss our
strategy for spectroscopic identification of this sample, and in Section \ref{sec:0804} we present the first new AM CVn star found in this way. We discuss the newly discovered system in Section \ref{sec:discussion} and conclude in Section \ref{sec:conclusion}.

\section{Sample selection}
\label{sec:sample}

\begin{figure}
\includegraphics[width=\columnwidth]{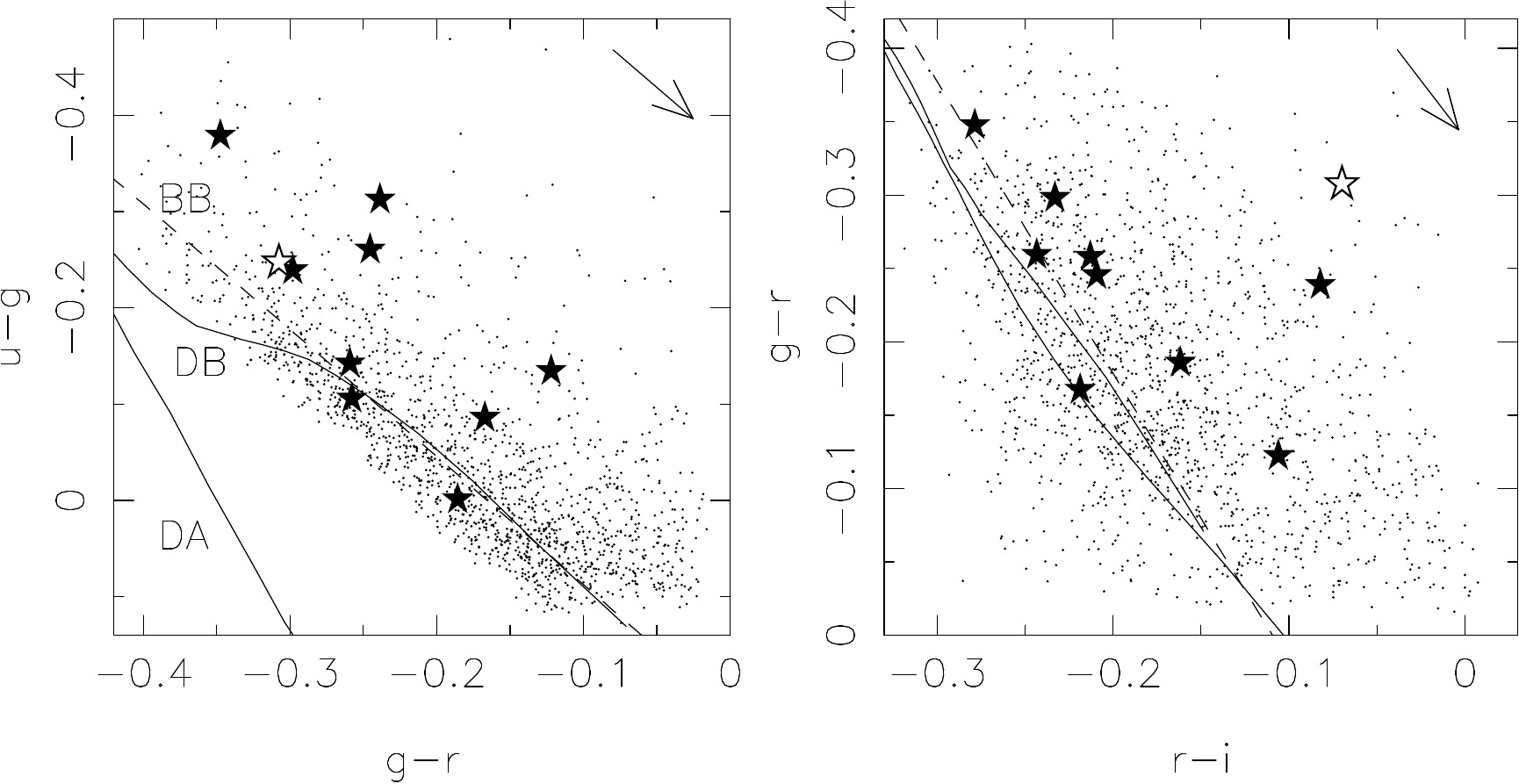}
\caption{Colours of the known AM CVn stars in the SDSS-DR6 (stars), together with our sample of candidates (dots). The dashed line (BB) shows the blackbody cooling track, and the solid lines (DA, DB) show model sequences for hydrogen and helium atmosphere white dwarfs, respectively. \obj\ is shown as the open star symbol. Arrows indicate reddening vectors for an extinction $A(g)=0.2$.}
\label{fig:colours}
\end{figure}

As shown in \citet{roelofspop}, the known AM CVn stars span a
range in $(u-g, g-r)$ colour space that lies substantially and slightly above the DA and DB white dwarf
cooling sequences, respectively. The DAs are much redder in $u-g$ due to the Balmer jump in their spectra, which is absent in the hydrogen-deficient AM CVns and DBs. See Figure \ref{fig:colours} in which we have plotted the $(u-g, g-r)$ and $(g-r, r-i)$ colour diagrams of the known AM CVn
stars, the colours of the AM CVn candidates selected here, and model colours of the cooling sequences of DA and DB white dwarfs of mass $\log g=8.0$. The latter were calculated from model spectra kindly provided by D. Koester.

All object colours are the
\emph{dereddened} colours from the SDSS-DR6, which were corrected for the full Galactic reddening according to \citet{schlegel}. The assumption
is that, at the high galactic latitudes spanned by the SDSS, almost all of the AM CVns will be further from the Galactic Plane than the scale height of dust
in the Galaxy. 
Models of the AM CVn
population \citep{nelemans,nelemans2004} show that the systems are expected to be several Gyr old on average, measured from the zero-age main sequence: similar to
old thin-disk main-sequence stars that are characterized by a Galactic
scale height $\sim$300 pc. Galactic dust, on the other hand, is known
to have a scale height of about 100 pc \citep{savagemathis}. Angular momentum losses due to the emission of gravitational waves will cause an AM CVn binary to evolve to an orbital period longer than 30 minutes in only $\sim$100\,Myr, counted from orbital period minimum. Combined with reasonable assumptions about the star formation history and the distribution of the delay times between star formation and the birth of AM CVn stars, this implies that $\sim$98\% of the local AM CVn population should have $P_\mathrm{orb}>30$\,min. In this `long' orbital period regime the AM CVns are expected \citep{bildsten} and observed \citep{thorstensen,roelofs,roelofsaw,roelofshst} to be dominated in the optical by an accreting white dwarf of typical temperature 10,000--20,000\,K, with a corresponding absolute magnitude of $M_g\approx10-13$. This implies that the bulk of the AM CVns (as well as hot, single white dwarfs) down to an apparent magnitude $g=20.5$ in the SDSS will be several 100\,pc, i.e.\ several dust scale heights, above the Galactic Plane. It also implies that, for the faintest systems, we are still covering the AM CVns out to about one scale height of the Galactic thin disk population ($\sim$300\,pc) if we include objects as faint as $g=20.5$. At fainter magnitudes, we anticipate that spectroscopic follow-up will become exceedingly difficult.

\begin{figure}
\includegraphics[width=\columnwidth]{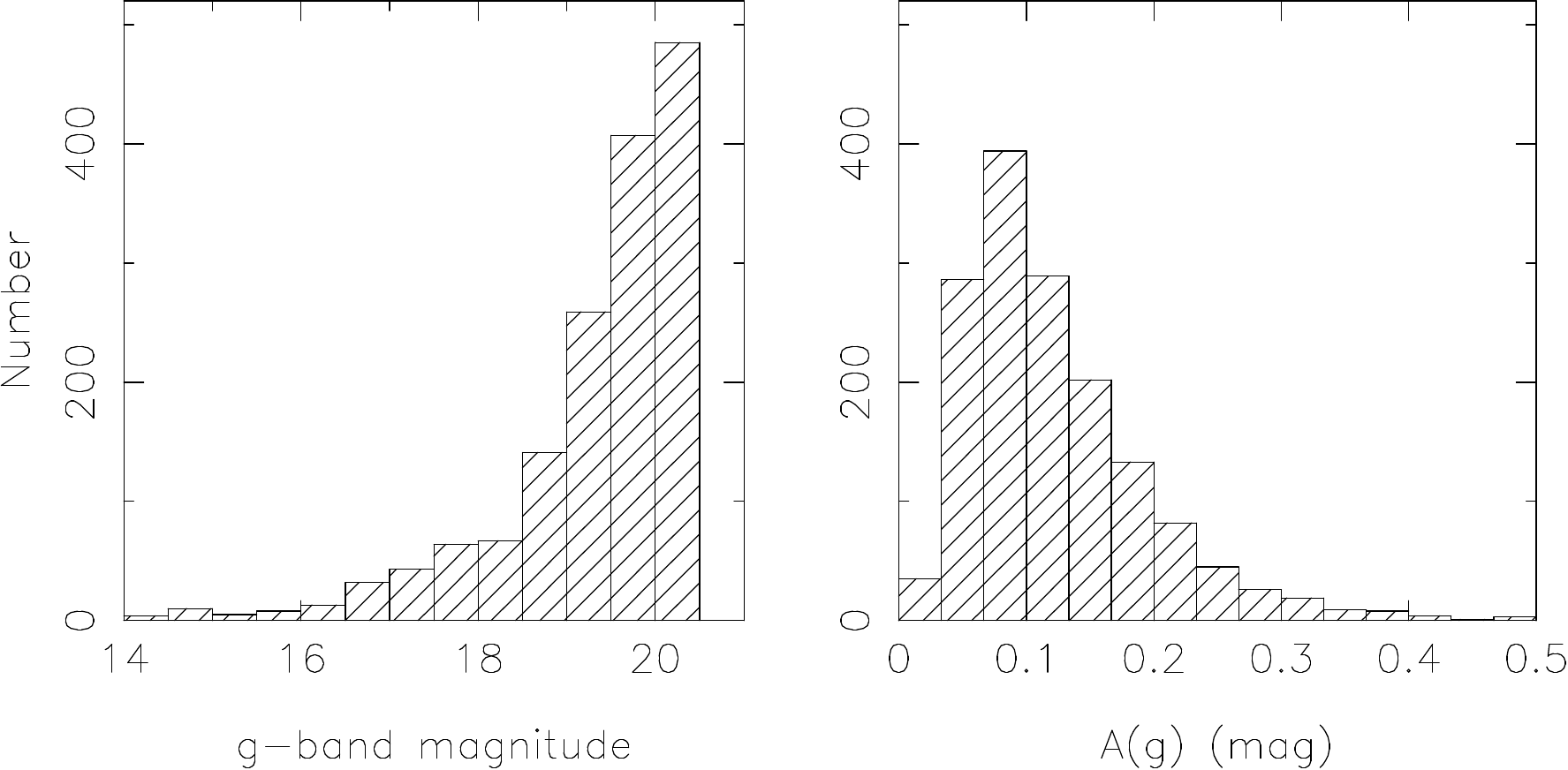}
\caption{Distributions of apparent $g$-band magnitude and Galactic $g$-band extinction $A(g)$ in our sample shown in Fig.~\ref{fig:colours}.}
\label{fig:samplestats}
\end{figure}

Guided by the colours of the known AM CVn stars, and attempting to reduce contamination by other sources  as much as possible (see below), we have queried
the SDSS-DR6 photometric database with the following relatively simple set of criteria (colours are dereddened):

\smallskip
\smallskip
\begin{tabular}{@{}r@{~~~}l@{}}
1.&Object is a point source\\
2.&Object is not saturated in any filter\\
3.&$g < 20.5$\\
4.&$u-g < \min\left[0.14, 1.35\left(g-r\right)+0.32\right] - \sigma_{u-g}$\\
5.&$-0.42+\sigma_{g-r} <g-r< 0.02-\sigma_{g-r}$\\
6.&$-0.33+\sigma_{r-i} <r-i< 0.03-\sigma_{r-i}$\\
\end{tabular}
\smallskip
\smallskip

\noindent where
\begin{equation}
\sigma_{u-g} = \sqrt{\sigma_u^2 + \sigma_g^2 + \sigma_{E(u-g)}^2}
\end{equation}
and similar for the other colours; we have assumed an uncertainty in the reddening of
\begin{equation}
\sigma_{E(u-g)}=0.5E(u-g)
\end{equation}
and the uncertainties $\sigma_{u,g,r,i}$ are the photometric errors on $u,g,r,i$ from the SDSS.
We take the uncertainty in the reddening to be a fraction of the reddening estimate, as suggested by \citet{schlegel}, except that we assume a larger fraction to further reduce the influence of sources in (relatively rare) fields with high reddening scattering into our colour selection. \citet{schlegel} derive an average scatter $\sigma_{E}=0.16E$, but this is for objects that are reddened by the full Galactic value; our sample includes Galactic sources at lower-than-full reddening which leads to a larger scatter. While the fractional error should thus be in the $0.16 < \sigma_E / E < 1$ range, the exact choice is arbitrary.

The resulting sample is plotted in Figure \ref{fig:colours}, and the distributions of $g$-band magnitude and extinction are shown in Figure \ref{fig:samplestats}. On the blue and sparsely populated
side in $(u-g, g-r)$ the colour selection includes the
10.3-minute orbital period AM CVn star ES Ceti \citep{wwescet}, the top-left star in both panels of Fig.\ \ref{fig:colours}, which is characterized by strong \mbox{He\,{\sc ii}} emission
lines and a very blue continuum. The bulk of the population, at orbital periods $>$30\,min, are expected to be cooler than ES Cet, since they are expected to have a much lower present-day accretion rate and have had a cooling time of $>$100\,Myr since the short-period, high accretion rate phase (see \citealt{bildsten}). The \emph{known} long-period systems are indeed observed to be redder in $u-g$, $g-r$ and $r-i$.

On the red side in $u-g$, care
has been taken to minimize the contribution from the DA and DB
white dwarf sequences by cutting off the sample parallel to the blackbody cooling track, as shown in Figure \ref{fig:colours}. The number of objects in the sample obviously increases rapidly with a redder cut-off on this side. While the exact number of candidates in the sample is arbitrary, the cut-off was chosen such that the number of objects remained manageable while still encompassing an estimated $\sim$90\% of the total population, based on the 9 known emission-line AM CVn stars in the SDSS-DR6 photometric database (plotted in Figure \ref{fig:colours}). It is important to note that the completeness of spectroscopic follow-up in the SDSS, i.e.\ the fraction of photometric objects that have an SDSS spectrum, \emph{increases} towards the more densely populated DA white dwarf locus in $(u-g,g-r)$ that is cut off from our sample, as shown in \citet{roelofspop}. The sample of 9 known AM CVns in the SDSS photometric database, of which six were discovered in the SDSS spectroscopic database, is thus actually slightly biased towards the red cut-off in $u-g$, making it unlikely that a large fraction of AM CVns fall outside this cut-off. It should be possible to more accurately judge and correct for the incompleteness of our sample \emph{a posteriori}, once the sample of spectroscopically identified AM CVns has been enlarged.

We removed known, previously identified objects (based on a \textsc{simbad}\footnote{\texttt{http://simbad.u-strasbg.fr/}} query) from the list of sources selected by the criteria above, which mostly affected objects at the bright end of the sample, and we checked the remaining objects by eye, removing obvious `mistakes' of the SDSS photometric data reduction pipeline. These were mostly compact star forming regions flagged as point sources in extensive starburst galaxies, as well as stars close to other very bright stars. Many of these were flagged `blended' \& `not deblended' in the photometric database; however we chose not to rely on those flags since a comparable number of seemingly fine single point sources were similarly flagged. Our final visual check weeded out a relatively small number of \vetoed\ sources from the sample.

Assuming a completeness of 90\%, our final sample of \sample\ candidates should
contain $\sim$40 new AM CVn stars, based on the six AM CVn stars found in the SDSS and the completeness of SDSS spectroscopy derived in \citet{roelofspop}. Spectroscopic identification of the sample should thus result in 1 AM CVn star per $\sim$40 objects, although the number will depend on the exact
characteristics of the population. Determining these characteristics is the aim of the study.

\section{Spectroscopic identification}
\label{sec:observing}

\begin{figure*}
\includegraphics[width=\textwidth]{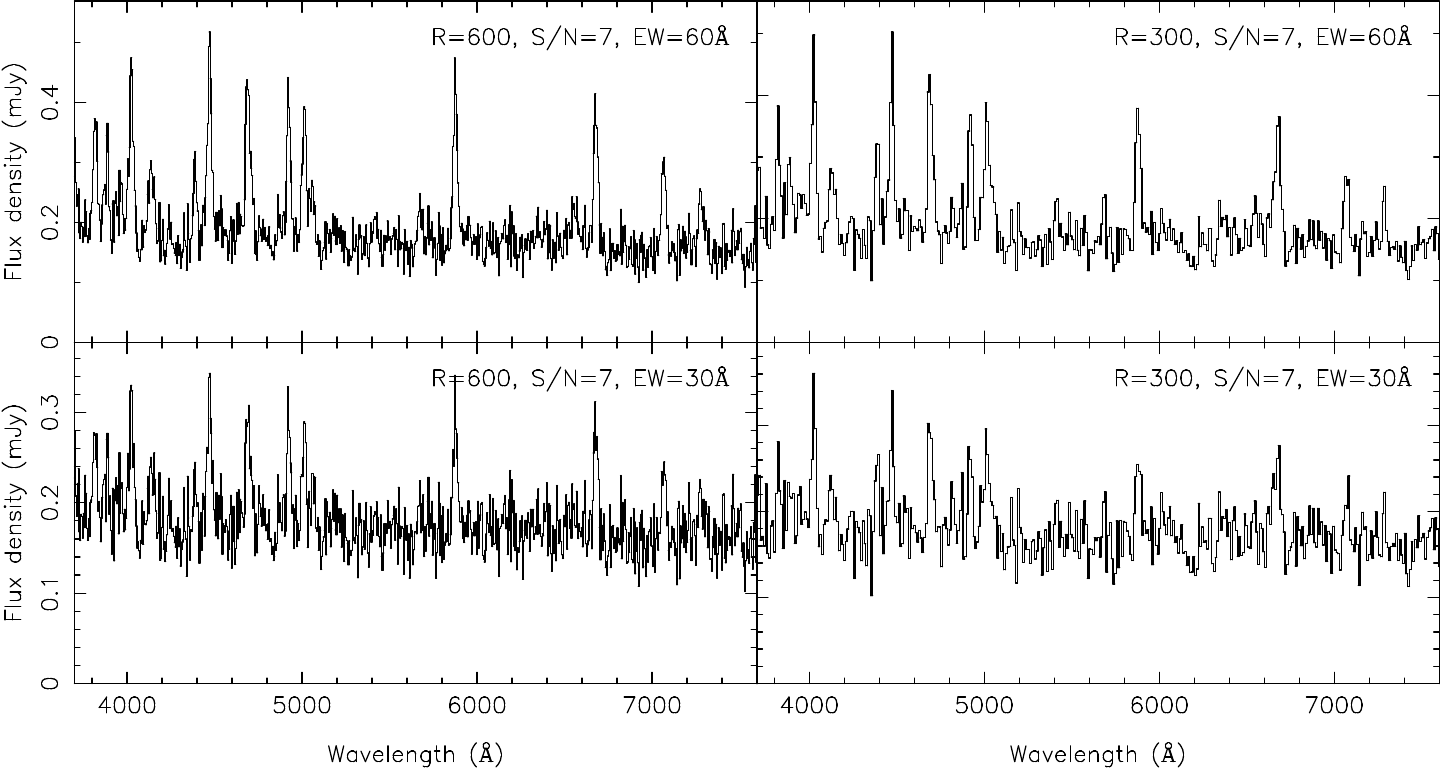}
\caption{Simulated low-resolution, low signal-to-noise ratio spectra of the new AM CVn star, \obj, to illustrate the requirements for spectroscopic identification of AM CVn stars. The spectral range is typical for a low-resolution set-up. Left and right panels show spectral resolutions $R=600$ and $R=300$, respectively, where each resolution element is sampled by two pixels. All spectra have a signal-to-noise ratio of 7 per pixel, or 10 per resolution element. In the lower panels the strengths of the lines have been reduced by a factor 2, to a \mbox{He\,{\sc i} 5876} equivalent width of 30\,\AA.}
\label{fig:simspec}
\end{figure*}

The known long-period systems are all characterized by
strong \mbox{He\,{\sc i}} emission lines in the optical, with \mbox{He\,{\sc i} 5876} equivalent widths ranging from 30\,\AA\ in V406 Hya and SDSS\,J1240$-$0159 ($P_\mathrm{orb}=34$ and 37 min, respectively; \citealt{roelofsaw}) to 90\,\AA\ in V396 Hya ($P_\mathrm{orb}=65.1$ min; \citealt{ruiz}). If we assume that the currently unknown population at $P_\mathrm{orb}>30$\,min has similar spectral characteristics (see \citealt{roelofspop} for a more detailed discussion), spectra with a resolution as low as $R=300$ and a signal-to-noise ratio (S/N) as low as 7 per pixel will suffice to positively identify all AM CVns among the candidates.

Figure \ref{fig:simspec} shows the spectrum of the new AM CVn star \obj\ (discussed below) degraded to resolutions of $R=600$ and $R=300$, and a S/N of 7 per pixel. It furthermore shows the same spectra again but with the equivalent widths of all spectral lines reduced by a factor 2, to make them more comparable to the AM CVn stars with the weakest lines of the known long-period systems. Each resolution element is sampled by exactly two pixels, giving a S/N of 10 per resolution element.

By considering the full optical spectrum, AM CVn stars like the ones we know today will clearly stand out even in the lowest-quality spectra shown. Even systems with weaker lines will still show a clear hint of helium emission, that can be confirmed from a second, higher-quality spectrum. If one only requires spectra of such low resolution and S/N, it becomes feasible to observe the entire sample of \sample\ objects with relatively small telescopes, despite the fact that the sample goes deep.

The question is, of course, whether there exist systems that, for some reason, have much weaker emission lines and may go undetected using this observing strategy. These cannot be ruled out given that the known systems have largely been selected on their emission lines. However, we consider it unlikely that these represent a large fraction of the population. There is a clear trend of increasing equivalent width of the emission lines with orbital period in the known AM CVns (Fig.~10 in \citealt{grootuvex}), presumably as a result of the accreting white dwarf cooling towards longer orbital periods, leading to a drop in the continuum flux against which the emission lines from the accretion disc are measured (a possible additional effect is the expected increase in the physical size of the disc towards longer periods). This suggests that systems with weaker emission lines will predominantly have short orbital periods, $P_\mathrm{orb}\lesssim30$\,min. As mentioned before, such short-period systems should represent only a small fraction of the total population. As in the case of the colour selection of our sample, the incompleteness of the spectroscopic identification of AM CVns can be judged more accurately \emph{a posteriori}, when all the actual spectra have been taken, and when it is clearer what the distribution in emission line strength is.

Spectra have so far been obtained of \done\ of the sample, mostly at the bright end,
in the Spring of 2008 using several telescopes, predominantly the 1.5-m Tillinghast telescope at Mt.\ Hopkins, Arizona equipped with the FAST spectrograph, and the 2.5-m Isaac Newton
Telescope on La Palma with the Intermediate Dispersion Spectrograph. One of the objects observed to date, \object\ (hereafter \obj), showed the anticipated spectrum of helium emission lines but no hydrogen, and will be discussed in detail in the next section.

\section{The first confirmed AM CVn star}
\label{sec:0804}

\subsection{Observations and data reduction}

\begin{table}
\begin{center}
\begin{tabular}{l l l l l}
\hline
Date       &UT             &Instrument   &Exposures\\
           &               &             &(exp.\ time)\\
\hline
\hline
2008/03/19 &00:55--03:03   &IMACS $f/4$  &46 (150\,s)\\
2008/04/17 &23:25--01:02   &MagE         &29 (180\,s)\\
2008/04/18 &23:19--01:05   &MagE         &32 (180\,s)\\
2008/04/19 &23:21--01:16   &MagE         &35 (180\,s)\\
\hline
\end{tabular}
\caption{Summary of our Magellan observations of \obj. Weather in all cases was good, with an average seeing of 0.6$''$.}
\label{table:observations}
\end{center}
\end{table}

\obj, a $g=18.2$ object in our sample (see also Fig.~\ref{fig:colours}), was found to be a strong helium-emission-line object based on a spectrum obtained 25 February 2008 with the 2.5-m Isaac Newton Telescope, La Palma, and the Intermediate Dispersion Spectrograph. Unusually strong lines of \mbox{He\,{\sc ii}} were noted, although contamination with the hydrogen Balmer series could not be ruled out for some of them.

In order to look for spectral line variations that might indicate orbital motion in the candidate AM CVn star, we obtained phase-resolved spectroscopy on the Magellan telescopes at Las Campanas Observatory, Chile, as listed in Table \ref{table:observations}. We initially used the IMACS $f/4$ camera with the 600 lines/mm grating to obtain a total of 46 consecutive spectra of 150\,s exposure time each on March 19, 2008. The spectral resolution was 3.0\,\AA. Wavelength calibration was obtained from helium--neon--argon comparison lamp exposures, with 0.2\,\AA\ root-mean-square residuals.

Following the tentative detection of spectroscopic variability on a $\sim$40-min timescale, we obtained a longer series of 96 spectra of 180\,s each on 17, 18 and 19 April 2008 using the new Magellan Echellette (MagE) spectrograph, covering the entire optical range from the atmospheric cut-off in the far-blue to approximately 9500\,\AA\ in the far-red, at a resolution $R\sim5000$ or about 1.2\,\AA. Detector read-out time between exposures was 23 seconds in `fast' mode, unbinned. Thorium--argon arc spectra were taken before and after the science exposures on each night, and dispersion solutions were obtained from third-order polynomial fits to the reference lines in the extracted echelle orders, leaving 0.05\,\AA\ root-mean-square residuals. The dispersion solutions for the science exposures were interpolated from the comparison arc solutions; arc drift was observed to be negligible over the time-span of our observations. Series of short and long incandescent lamp exposures, providing median-filtered flat field images for the red and blue sides of the CCD respectively, were combined into a single normalised flat field. The echelle orders were flux calibrated using comparison spectra of the spectrophotometric standard star HD\,60753. At wavelengths $>5000$\,\AA, we further corrected for telluric absorption lines using comparison spectra of the DA white dwarf standard LTT\,3218. Correction for atmospheric extinction and wavelength-dependent slit-losses remained imperfect since the standard star had to be observed at a lower airmass than \obj; this could affect the overall slope of the spectrum. The corrected echelle orders were merged while weighting pixels in overlapping orders by number of counts, and finally the wavelength scale was transformed to the heliocentric rest-frame.

\subsection{Results}

\subsubsection{Average spectrum}

\begin{figure*}
\includegraphics[width=\textwidth]{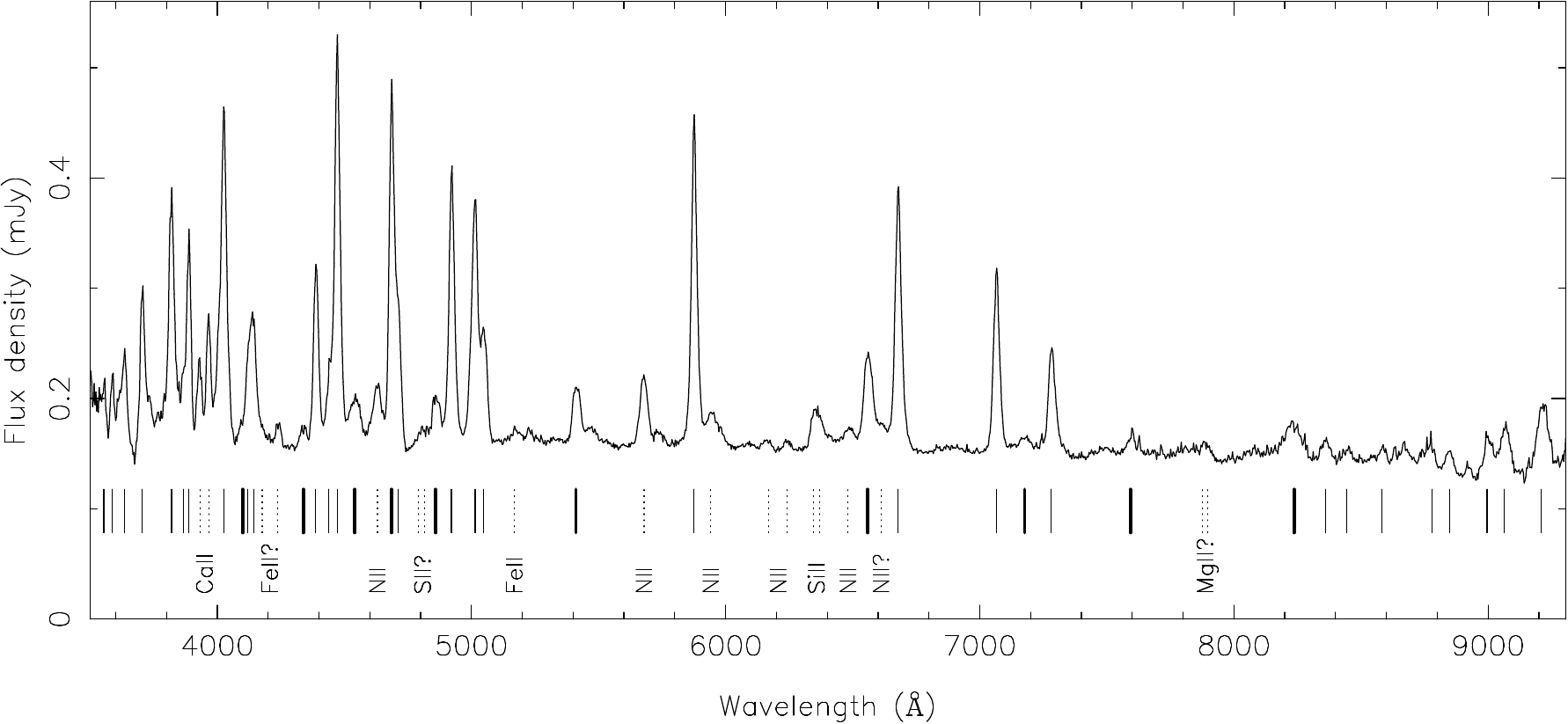}
\caption{Average spectrum of \obj. Thin and thick solid lines indicate neutral and ionised helium lines, respectively, while dotted lines indicate metal lines as labelled. See Figure \ref{fig:lines} for a detailed view of the shapes of the emission lines.}
\label{fig:spectrum}
\end{figure*}

Figure \ref{fig:spectrum} shows the average spectrum of \obj. It exhibits a plethora of emission lines, mainly of helium. Emission lines close to the hydrogen Balmer series are found to be \mbox{He\,{\sc ii}}, based on their central wavelengths and strengths relative to the \mbox{He\,{\sc ii}} lines at 5411 and 4541\,\AA\ that are not coincident with hydrogen. The emission lines at the far-red end of the spectrum do not coincide with the hydrogen Paschen series, and it thus appears there is no evidence for the presence of hydrogen.

A series of lines observed at 4630, 5679, 5941, 6170, 6242, 6482 and possibly 6610\,\AA\ are identified with \mbox{N\,{\sc ii}}. These, together with the observed \mbox{He\,{\sc ii}} lines, indicate temperatures of $\sim$20,000--27,000\,K in at least part of the accretion flow. However, several strong carbon and oxygen lines that would be expected at these temperatures are missing entirely from the spectrum, most notably \mbox{C\,{\sc ii}} at 4267, 6785, 7119 and 7236\,\AA, and either \mbox{O\,{\sc i}} 7774 or \mbox{O\,{\sc ii}} 4070-4077, 4591\,\AA. Highly non-solar abundance ratios of N/O$\,\gtrsim\,$10 and N/C$\,>\,$10 by number are needed to sufficiently suppress the carbon and oxygen lines relative to nitrogen regardless of the exact temperature structure, assuming a simple spectral model of a gas mixture in local thermodynamic equilibrium (LTE): the same model that was used to derive a similar (yet stronger) limit of N/C$\,>\,$100 for the 46-min binary GP Com (\citealt{trm91}, as updated by \citealt{nelemansco}). This strongly non-solar abundance pattern has been seen as evidence for CNO-processing in GP Com, with the added complication that the apparent absence of metal lines such as Si in that system suggests a metal-poor initial composition, which would require that the observed N have been dredged up and transferred during the prior AGB phase of what is now the accreting white dwarf. The \mbox{Si\,{\sc ii}} 6347 \& 6371 lines that were `missing' in GP Com are clearly observed in \obj\ (and have been observed in spectra of several other AM CVns: \citealt{grootcperi,roelofs,roelofsaw}), alleviating the problem of the origin of all the nitrogen for \obj\ (but not for GP Com).

The strong emission-line complexes of \emph{neutral} nitrogen seen in GP Com are absent, or at the very least least much weaker, in \obj\ despite the clear presence of nitrogen in its ionised form (see Table \ref{table:ews}).

\begin{figure}
\includegraphics[width=\columnwidth]{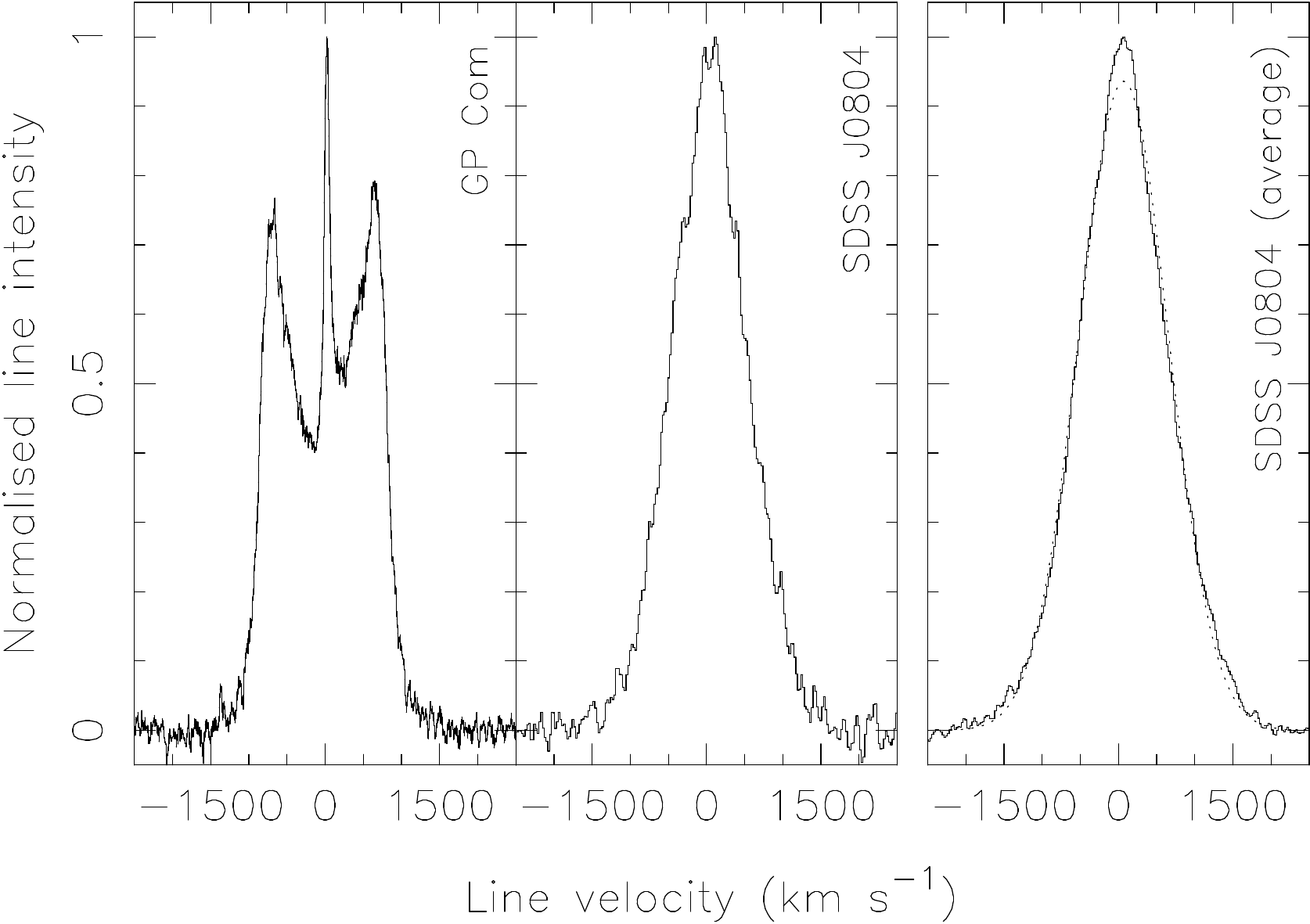}
\caption{Profile of the \mbox{He\,{\sc i}} 7065 line in \obj\ compared to the more typical profile in GP Com (GP Com data from Steeghs et al.\ in preparation). The line profiles are well resolved, with resolution elements corresponding to approximately two bins. The right panel shows the average of the \mbox{He\,{\sc i}} 4387, 4471, 4921, 5875, 6678 and 7065 lines, and a Gaussian fit indicated by the dotted line.}
\label{fig:lines}
\end{figure}

The emission lines in \obj\ have a full-width at half-maximum of about 1,500 km/s and are approximately Gaussian in shape, except for a slight excess of flux at low velocities. See Figure \ref{fig:lines} for a close-up of the \mbox{He\,{\sc i}} 7065 line in comparison with the same line in GP Com. The latter has the more typical triple-peaked profile seen in AM CVn stars: a double-peaked profile from the accretion disc, similar to the profiles seen in non-magnetic, hydrogen-rich CVs, plus a very narrow spike unique to the AM CVns, which is thought to arise from (very close to) the surface of the accreting white dwarf and to be related to the accretion process (see \citealt{trm99,lmr}, and see \citealt{ruiz,roelofsaw,roelofs1552} for emission-line profiles in a number of other AM CVns). None of the emission-line profiles of \obj\ show any evidence for a classical double-peaked component from a disc, indicating a different geometry of the accretion flow.

Table \ref{table:ews} lists the widths and equivalent widths of most of the lines observed in the spectrum of \obj, and compares them to those observed in the 46-minute binary GP Com.

\begin{table}
\begin{center}
\begin{tabular}{l r r r}
\hline
Line                       &FWHM (km s$^{-1}$) &&EW ($-$\AA)\\[1ex]
\cline{3-4}\\[-2.ex]
                           &                   &\emph{SDSS\,J0804}       &\emph{GP Com}\\
\hline
\hline
\mbox{He\,{\sc ii}} 4340  &\gpm{2300}{300}&\gpm{3.3}{0.4}      &$-$\\
\mbox{He\,{\sc i}} 4387   &\gpm{1550}{\phantom{0}50}&\gpm{22.6}{1.0}     &\gpm{12.8}{0.3}\\
\mbox{He\,{\sc i}} 4471   &\gpm{1500}{100}&\gpm{55.0}{1.0}     &\gpm{39.9}{0.3}\\
\mbox{He\,{\sc ii}} 4541  &\gpm{2400}{150}&\gpm{8.0}{0.8}     &$-$\\
\mbox{N\,{\sc ii}} 4630   &\gpm{2200}{150}&\gpm{10.0}{1.0}     &\gpm{1.0}{0.3}\\
\mbox{He\,{\sc ii}} 4686  &\gpm{1500}{150}&\gpm{78.0}{2.0}     &\gpm{24.5}{0.3}\\
+ \mbox{He\,{\sc i}} 4713 &&                    &\\
\mbox{He\,{\sc ii}} 4860  &\gpm{1900}{100}&\gpm{8.0}{0.8}      &$-$\\
\mbox{He\,{\sc i}} 4921   &\gpm{1460}{\phantom{0}50}&\gpm{48.0}{1.0}     &\gpm{29.2}{0.3}\\
\mbox{He\,{\sc i}} 5015   &\gpm{1600}{150}&\gpm{62.0}{1.0}     &\gpm{47.8}{0.3}\\
+ \mbox{He\,{\sc i}} 5047 &&                    &\\
\mbox{He\,{\sc ii}} 5411  &\gpm{1900}{75}&\gpm{12.8}{0.5}     &$-$\\
\mbox{N\,{\sc ii}} 5680   &\gpm{1930}{50}&\gpm{15.0}{1.0}     &\gpm{3.3}{0.3}\\
\mbox{He\,{\sc i}} 5876   &\gpm{1250}{50}&\gpm{60.0}{1.0}     &\gpm{77.7}{0.3}\\
\mbox{N\,{\sc ii}} 5942   &$\cdots$&\gpm{9.0}{1.0}      &$-$\\
\mbox{N\,{\sc ii}} 6170   &$\cdots$&\gpm{2.2}{0.2}      &$-$\\
\mbox{N\,{\sc ii}} 6242   &$\cdots$&\gpm{2.3}{0.2}      &$-$\\
\mbox{N\,{\sc ii}} 6484   &$\cdots$&\gpm{4.5}{0.4}      &$-$\\
\mbox{He\,{\sc ii}} 6559  &\gpm{1700}{50}&\gpm{20.0}{1.0}     &$-$\\
\mbox{He\,{\sc i}} 6678   &\gpm{1240}{50}&\gpm{50.0}{2.0}     &\gpm{60.6}{0.3}\\
\mbox{He\,{\sc i}} 7065   &\gpm{1240}{50}&\gpm{34.8}{0.5}     &\gpm{54.3}{0.3}\\
\mbox{He\,{\sc ii}} 7177  &$\cdots$&\gpm{2.2}{0.4}     &$-$\\
\mbox{He\,{\sc i}} 7281   &\gpm{1280}{50}&\gpm{20.0}{1.0}     &\gpm{27.2}{0.4}\\
\mbox{N\,{\sc i}} 7440    &&$-$      &\gpm{14.1}{0.4}\\
\mbox{O\,{\sc i}} 7774    &&$-$                 &\gpm{5.0}{0.4}\\
\mbox{N\,{\sc i}} 8200    &$\cdots$&$^\dagger$\gpm{19.0}{2.0}     &\gpm{51.1}{0.5}\\
\mbox{N\,{\sc i}} 8700    &$\cdots$&$^\dagger$\gpm{15.0}{2.0}     &\gpm{104.6}{0.6}\\
\hline
\mbox{Ca\,{\sc ii}} 3933  &\gpm{1400}{100}&\gpm{8.0}{0.8}      &$-$\\
\mbox{Si\,{\sc ii}} 6347  &$\cdots$&\gpm{7.0}{0.7}      &$-$\\
+ \mbox{Si\,{\sc ii}} 6371 &&                    &\\
\hline
\end{tabular}
\caption{Equivalent width (EW) and Gaussian full-width at half-maximum (FWHM) for the most prominent spectral lines in \obj. Estimated errors are mainly due to uncertainty in the continuum level. The last column shows GP Com data from \citet{trm91}. Non-detections, marked `$-$', imply an EW of $0.0\pm0.3$. Values marked `$\cdots$' could not be measured reliably. $^\dagger$Uncertain identification; EW of \mbox{N\,{\sc i}} component likely lower (and possibly zero).}
\label{table:ews}
\end{center}
\end{table}

\subsubsection{Spectral line variability}

\begin{figure}
\includegraphics[width=\columnwidth]{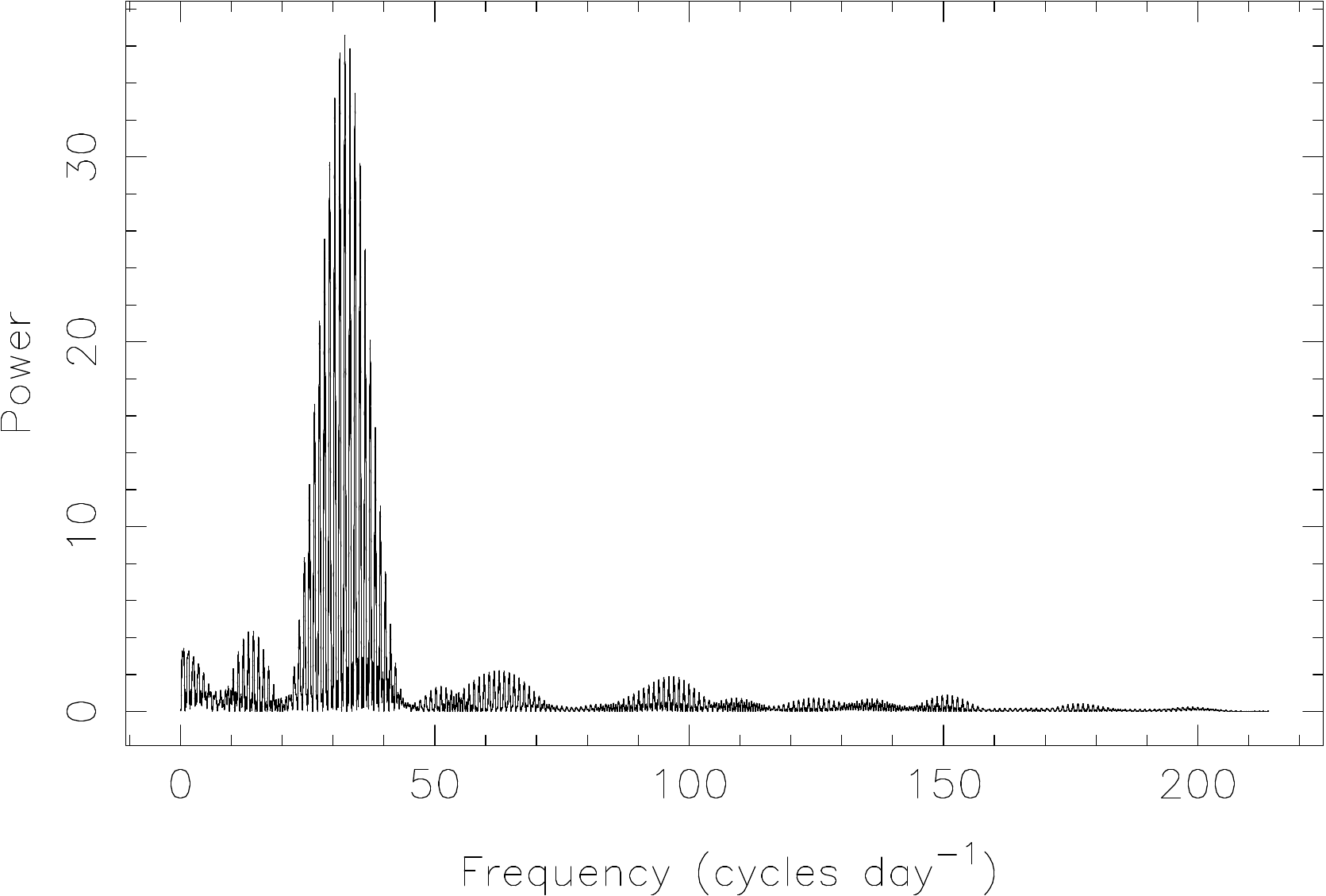}
\caption{Lomb--Scargle periodogram of the flux ratios of the red and blue wings of the emission lines used for Figure \ref{fig:trail}.}
\label{fig:scargle}
\end{figure}
\begin{figure}
\includegraphics[width=\columnwidth]{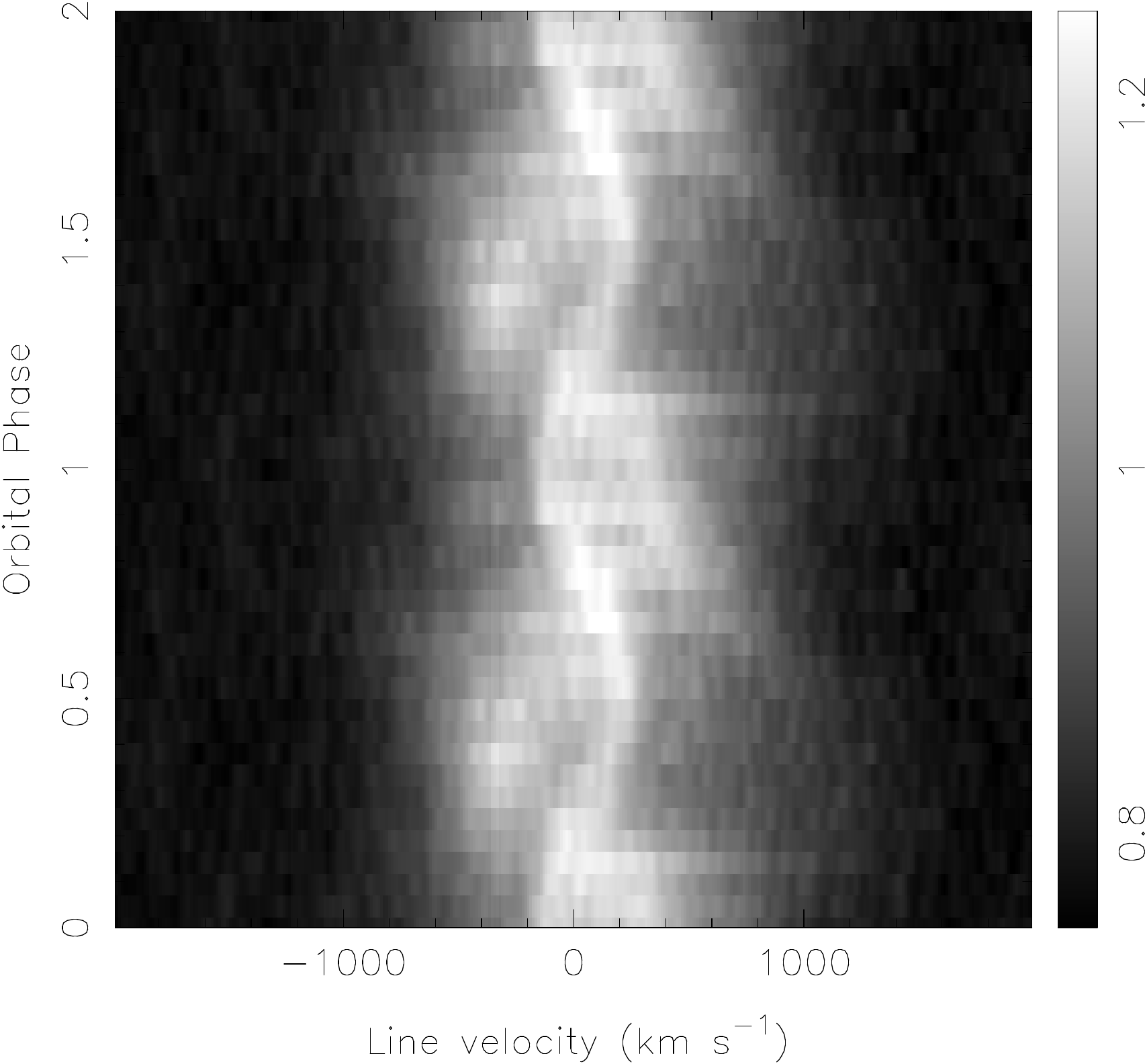}
\caption{Trailed spectrum of \obj, composite of the \mbox{He\,{\sc i}} and \mbox{Si\,{\sc ii}} lines between 5000--7400\,\AA. A sinusoidal `S-wave' variation is seen on a period of \porb. Higher-velocity emission, up to 1500 km/s, is seen to vary on the same period. The grey-scale indicates the relative flux densities.}
\label{fig:trail}
\end{figure}

Variability in the spectral lines, typically in the form of a sinusoidal `S-wave' in spectroscopic time-series, is observed in most of the AM CVn stars and all of the long-period, emission-line systems \citep{nather, ruiz, roelofs, roelofsaw, roelofs1552}. Following the procedure of \citet{nather} we compute a Lomb--Scargle periodogram of the ratios of fluxes in the red and blue wings of the emission lines of \obj. See Figure \ref {fig:scargle}. A strong variability peak is seen centered on 32.4 cycles per day, or 44.5 minutes. The short observing baseline and equal spacing of the individual blocks of spectra create strong aliases at 1 cycle/day intervals, but regardless of the spectral lines or velocity limits chosen for the red-wing/blue-wing periodogram, the 44.5-minute period comes out as the strongest peak.

Since the red-wing/blue-wing method does not optimally utilize the variability information contained in the spectroscopic time-series, we also do a `double-blind' (but subjective) test to judge by eye at which folding period, in the range 20--120 minutes, the strongest variability pattern occurs. This yields $P_\mathrm{fold}=44.5\pm0.1$\,min, a perfect match to the period obtained from the Lomb--Scargle periodogram. The corresponding trailed spectrum is shown in Figure \ref{fig:trail}. It shows a clear S-wave with a velocity amplitude of $160\pm10$\,km/s, as well as asymmetric higher-velocity emission extending to about 1,500\,km/s on the red side. Based on the sinusoidal S-wave feature we propose that $P_\mathrm{fold}=P_\mathrm{orb}=44.5\pm0.1$\,min.

A similar picture is seen at the $\pm1$ cycle/day aliases of this period, as expected, while for the $\pm2$ cycles/day (and more distant) aliases the pattern fades. No variability is seen on periods other than aliases of the main signal. We consider the $\pm1$ cycle/day aliases possible, and the more distant aliases unlikely, to represent the actual orbital period. The time coverage of the present data is insufficient to track the S-wave from one night to the next and break the aliases, but the data suffice to firmly establish the ultracompact nature of the binary, thereby confirming its membership of the AM CVn family.

With the S-wave period in hand, we revisit our IMACS data obtained one month earlier. No S-wave is seen in any of the lines or combination of lines, nor for any alias of the S-wave period. Although the IMACS data have a slightly lower resolution, lower signal-to-noise ratio, and shorter baseline, we would have expected to see a sign of the S-wave in the corresponding trailed spectrum if it were as strong as in the MagE spectra. We conclude that the pattern of orbital-period variations must change on timescales of one month.

\section{Discussion}
\label{sec:discussion}

The first new AM CVn star from our survey of candidates from the SDSS poses an interesting optical spectrum: devoid of hydrogen, but rich in nitrogen, with strikingly strong lines of ionised nitrogen and helium. The high nitrogen abundance puts constraints on the prior evolution of the donor star, in particular the amount of helium burning that may have taken place. 
Detailed helium star models for the donor stars in AM CVns predict typical surface abundance ratios N/C\,=\,0.001$-$1 at orbital periods $>$30\,min \citep{yungelson}, while for white dwarf donor stars one expects N/C\,$\approx$\,0.1 (if the CNO cycle never operated during the donor's hydrogen-burning phase) or N/C\,$\gg$\,1 (if the CNO cycle has converted most of the primordial C and O into N). This suggests that the donor star in \obj\ has not had a phase of helium burning, implying that the helium-star model for the formation of AM CVn stars (e.g.\ \citealt{ibentutukov}) does not apply to \obj. The lower limits of the observed N/C and N/O abundances could be \emph{marginally} consistent with the lowest-possible amount of helium burning in the models of \citet{yungelson}, corresponding to a helium star that is already almost filling its Roche lobe upon reaching the zero-age helium main sequence, but this requires fine-tuning. The `evolved-CV' model for the formation of AM CVn stars \citep{podsi} predicts essentially the same CNO abundances as the helium white dwarf model, and is thus compatible as far as the observed CNO abundances are concerned.

While the LTE emission line model used to derive the relative CNO abundances is rather simple, it seems unlikely that a more detailed treatment including non-LTE effects could succeed in suppressing all the expected lines of C and O while reproducing the observed lines of N, without an actual nitrogen overabundance. It would however be worthwhile to try to apply such models (e.g.\ \citealt{nagel}). An alternative hypothesis could be that sedimentation of C and O in the outer layers of the donor star precludes these elements from showing up in the spectrum, or that we are seeing the previously sedimented N in what is now just the inner few $0.01M_\odot$ of the original donor. This mechanism is unlikely to work, however, since none of these elements should sink into a degenerate helium donor star, given the identical mass-to-charge ratios of the C, N, O and He nuclei involved.

A large nitrogen abundance is far from unique among the AM CVn stars, rather the opposite: anomalously large N/C line ratios in the far-UV spectrum of V396 Hya were noted by \citet{gaensicke}, while the X-ray spectra of several others were best modelled with CNO-cycled or even higher nitrogen abundances \citep{ramsayamcvns,ramsayv396}. Combined with the results of \citet{yungelson}, this would suggest that most of the known AM CVns do not harbour formerly helium-burning donor stars, even though the large observed luminosities of the short-period AM CVns are much more readily explained if one does assume such donors \citep{roelofsamcvn,roelofshst}.

Further information about \obj\ may be obtained from the profiles of its emission lines. Their unique single-peakedness is not simply an inclination effect, since the lines are very broad compared to those in other emission-line AM CVns that still clearly show a double-peaked accretion disc profile (e.g., \citealt{roelofs}). The most likely explanation, also in view of the large \mbox{N\,{\sc ii}}/\mbox{N\,{\sc i}} and \mbox{He\,{\sc ii}}/\mbox{He\,{\sc i}} line ratios, is that the system is magnetic: broad, single-peaked, almost Gaussian emission lines with strong series of ionised helium are characteristic of magnetic CVs and the spectrum of \obj\ is a perfect match, save for the absence of hydrogen (e.g.\ \citealt{schwartz}). Optical polarimetry should be pursued to confirm the magnetic nature of the system, and near-infrared spectroscopy might reveal cyclotron emission features as further evidence for the presence of strong magnetic fields.

A final, intriguing question is whether we should not have found more than one AM CVn star amongst the candidates observed so far. The hit-rate does appear to be below expectations (Section \ref{sec:sample}), but may at this point simply suffer from low-number statistics and the fact that we have so far only targeted the bright end of the sample, which means that we have probably observed a different population of objects than the average of the entire sample. Once the entire sample has been completed, a full analysis as described in \citet{roelofspop} will have to be done, which will give a revised and more accurate space density for the Galactic AM CVn population.

\section{Conclusion}
\label{sec:conclusion}

We have described a spectroscopic survey aimed at identifying all of the $\sim$40 expected emission-line AM CVn stars in an estimated 90\% complete sample of \sample\ candidates, drawn from a colour selection of the Sloan Digital Sky Survey, Data Release 6, down to $g=20.5$. The aim is to greatly increase the size and completeness of the sample of AM CVn stars from the SDSS, allowing for a much more accurate assessment of the total Galactic population.

We have presented time-resolved spectroscopy confirming the first new AM CVn star from this survey, \object, and measured an orbital period of \porb, or possibly one of its nearest daily aliases. The single-peaked emission lines as well as the strong lines of ionised helium, both unique among the long-period AM CVns known to date, are very much reminiscent of the magnetic systems among the hydrogen-rich Cataclysmic Variables, and we suggest that \obj\ may be the first ultracompact version of such a system.

The strong lines of nitrogen but non-detections of expected lines of carbon and oxygen suggest that the donor star in \obj\ contains highly non-solar abundance ratios of these elements. We have derived, from a simple LTE model, lower limits of N/C$\,>\,$10 and N/O$\,\gtrsim\,$10 by number. This indicates that the donor star has burned hydrogen via the CNO cycle, while it is very unlikely that it has had a subsequent helium-burning phase, effectively ruling out the `helium-star' formation scenario for this system.

\section*{Acknowledgments}

GHAR is supported by NWO Rubicon grant 680.50.0610 to G.H.A. Roelofs. DS acknowledges an STFC Advanced Fellowship. GN is supported by the Netherlands Organization for Scientific Research. Based on data taken at the Magellan Telescopes at Las Campanas Observatory, Chile, and at the Isaac Newton Telescope in the Observatorio del Roque de los Muchachos of the Instituto de Astrof\'isica de Canarias, Spain. This study makes use of the Sloan Digital Sky Survey; see \texttt{http://www.sdss.org/collaboration/credits.html} for the full acknowledgment. Special thanks to P. Berlind and the 1.5-m telescope crew at F.L. Whipple Observatory (a facility of the Smithsonian Institution) for their valuable contribution, and to L. Yungelson for helpful discussion. Figure \ref{fig:colours} makes use of white dwarf model spectra kindly provided by D. Koester.

{}

\end{document}